# Design & Implementation of Automatic Machine Condition Monitoring and Maintenance System in Limited Resource Situations

Abu Hanif Md. Ripon[1*], Muhammad Ahsan Ullah[2,] Arindam Kumar Paul[3], Md. Mortaza Morshed[4]

[1,2]Department of Electrical and Electronic Engineering
Chittagong University of Engineering and Technology
[3]Research Assistant, Mathematics Discipline, Khulna University, Khulna – 9208.
[4]Deputy Inspector General, Department of Inspection for Factories and Establishment, Ministry of Labor and Employment, Government of the People's Republic of Bangladesh.
Email: abuhanifmd.ripon@gmail.com, ahasan@cuet.ac.bd, arindam017@gmail.com, morshedmortaza@gmail.com

## List of Abbreviations

| Abbreviations | Definitions |
| --- | --- |
| FDD | Fault Detection and Diagnosis |
| SVM | Support Vector Machine |
| DSP | Digital Signal Processing |
| NN | Neural Network |
| ML | ML Machine learning |
| AI | Artificial Intelligence |
| PCA | Principal Component Analysis |
| OSVM | Optimizable Support Vector Machine |
| CSV | Comma Separated Value |
| VAE | Variational Autoencoder |
| LDA | Linear Discriminant Analysis FDD |
| AR | Autoregressive |
| ARX | Autoregressive with EXogenous Input |
| ARMA | AutoRegressive Moving Average |
| ARARX | AutoRegressive with eXogenous input |
| ARMAX | AutoRegressive Moving Average with eXogenous input |
| DTW | Dynamic Time Warping |
| TDDP | Time-domain Dimensional Parameters |

**Abstract**
In the era of the fourth industrial revolution, it is essential to automate fault detection and diagnosis of machineries so that a warning system can be developed that will help to take an appropriate action before any catastrophic damage. Some machines health monitoring systems are used globally but they are expensive and need trained personnel to operate and analyse. Predictive maintenance and occupational health and safety culture are not available due to inadequate infrastructure, lack of skilled manpower, financial crisis, and others in developing countries. Starting from developing a cost-effective data acquisition system for collecting fault data in this study, the effect of limited data and resources has been investigated while automating the process. To solve this problem, A feature engineering and data reduction method has been developed combining the concepts from wavelets, differential calculus, and signal processing.  Finally, for automating the whole process, all the necessary theoretical and practical considerations to develop a predictive model have been proposed. For validation of the system performance, several faults have been created in the laboratory for training and testing. The proposed method outperformed the conventional method while testing. SVM and NN were proposed for the prediction purpose because of their high predicting accuracy greater than 95% during training and 100% during testing the new samples.  In this study, the combination of the simple algorithm with a rule-based system instead of a data-intensive system turned out to be hybridization by validating with collected data. The outcome of this research can be instantly applied to small and medium-sized industries for finding other issues and developing accordingly. As one of the foundational studies in automatic FDD, the findings and procedure of this study can lead others to extend, generalize, or add other dimensions to FDD automation.

# 1 Introduction

In modern automated industries, the early and efficient detection of faults is of utmost importance for the proper functioning of physical systems. Rotating machinery plays a crucial role in various industries, and machine condition monitoring and fault diagnosis have gained significant attention due to their potential benefits, including reduced maintenance costs, increased productivity, and improved machine availability. However, relying on traditional manual inspections or periodic maintenance often fails to identify underlying faults early, leading to unexpected breakdowns, higher repair expenses, and extended downtime [1]. Vibrations, which emerge at the onset of issues, offer valuable insights into machine health [2]. Analyzing vibration patterns can accurately pinpoint problems, but this requires expertise. Unfortunately, in many Bangladeshi industries, the lack of effective early fault detection processes poses challenges. To address this, Bangladesh needs to cultivate a skilled workforce capable of assessing machine health. Currently, the absence of cost-effective local technology and equipment for this purpose necessitates the importation of high-value condition monitoring devices from abroad [3]. In addressing the challenges of early fault detection in machinery, there is a notable focus on condition-monitoring techniques. This involves the development of a locally engineered data acquisition system tailored for capturing acceleration and vibration data from rotating machines. The system utilizes an ESP32 microcontroller and a gyro acceleration sensor to attain precise data, offering customization of data acquisition duration and yielding an impressive 1600 data points per second. This enables in-depth analysis and comprehensive insights into machine behaviour. The research also employs a robust data management strategy to ensure secure temporary storage and seamless transfer of collected data in a widely supported CSV format, ensuring data integrity and compatibility with various analysis tools. Additionally, to address the lack of human resources, Artificial Intelligence (AI) technology is suggested as a viable solution. Research demonstrates that AI can accurately diagnose machine problems by analyzing vibration signals, with SVMs (SVM) outperforming medium-class NNs in terms of accuracy. Advanced AI technology enhances the efficiency and accuracy of problem detection. The vibration signal analysis involves three key steps: data acquisition, signal processing, and fault classification. The research incorporates an optimizable SVM (OSVM) for fault detection, known for its fast classification, adept handling of non-linear data, and global optimum provision. Furthermore, Principal Component Analysis (PCA) is utilized to reduce the complexity of high-dimensional data, projecting it onto a lower-dimensional subspace while retaining essential information, making it valuable for visualization, compression, and feature extraction [2], [3]. Developing a locally engineered data acquisition system and using the collected data to fuel AI models is of immense importance to industries. It offers greater control over machine fault detection processes and promotes the widespread adoption of advanced condition monitoring practices. This research specifically concentrates on leveraging AI to detect machine problems by gathering data through cost-effective indigenous technology [3]. The goal is to introduce an appealing and innovative approach to identify machine faults within Bangladesh's industrial landscape accurately. Various methods, including DSP-based, machine learning-based, model-based techniques, acoustic emission techniques, and a combination of DSP and ML methods, have been employed to accomplish this objective [4], [5].

## 2  Composition of Methods and Concepts

The necessity for extensive and varied datasets to teach and verify models presents a notable challenge when implementing FDD in LDCs [13]. The complexity arises from the difficulty of adapting models to diverse operating conditions, environments, and devices due to limited research, development, and expert manpower [13]. The overlooked challenge of interpreting results and explaining model logic further complicates matters [5]. Moreover, the absence of standardized evaluation methods hinders comparing model performance [13].

To address these limitations, previously suggested methods include:
Utilizing data augmentation, transfer learning, or domain adaptation techniques to expand datasets and enhance model building [13].
Employing explainable artificial intelligence (XAI) methods to improve model transparency and interpretability [13].

In resource-constrained environments, numerous challenges are faced by automatic Fault Detection and Diagnosis (FDD) algorithms. A significant issue arises from the insufficiency of data, where these algorithms heavily rely on historical data for training and pattern recognition. The accuracy and reliability of FDD models are compromised in settings with limited resources due to the absence of robust data collection infrastructure, coupled with data quality issues and constraints in storage capacity.

Furthermore, difficulties emerge in the deployment and maintenance of FDD systems due to the scarcity of skilled manpower. The necessary expertise for algorithm development, model training, and system integration is often lacking in environments with limited resources. The interpretation of FDD results becomes a additional challenge, requiring domain knowledge and analytical skills that may be insufficient.

Infrastructure limitations further contribute to impeding the effectiveness of FDD algorithms. The computational intensity of these algorithms necessitates substantial hardware and software resources, which are often unavailable in resource-constrained settings. Additionally, the challenges posed by unreliable or limited communication networks hinder the transmission of data and remote monitoring capabilities of FDD systems. To address these challenges, strategic approaches such as the utilization of simplified algorithms, integration of knowledge-based systems, and collaboration with experts for remote assistance are crucial for enhancing the adaptability and resilience of FDD algorithms in resource-limited settings.

The study design and determination of the sequence and components of the research work were heavily reliant on subject area knowledge, given the limitations of resources and skilled manpower in lower-middle-income countries. The process of conducting experiments and making choices, both of which can be expensive and time-consuming, was guided by mathematical concepts and previous research findings. For feature extraction, dataset reduction, and model selection, we utilized mathematical concepts and previous research findings. The combination of PCA and SVM, shown to be effective in similar cases in previous research, was found not to be suitable for this particular dataset. In response to this observation, the effectiveness of wavelets was recognized, leading to the decision to use them for denoising the entire dataset. This choice allowed for the removal of insignificant noise without compromising valuable information. A unique and effective approach was employed for addressing the complexities and patterns within the dataset, involving the calculation of gradients of axis values for the vibration data. This approach aligns with the strengths of dynamical systems analysis and differential equations, which capture the rate of change over time, providing a more nuanced representation than ordinary systems of equations. Gradients and wavelet denoising often yield features that benefit from non-linear kernels such as Gaussian kernels, as they can effectively capture complex relationships within the data. A support vector machine (SVM) algorithm with a predefined Gaussian kernel was selected due to its ability to model smooth decision boundaries, well-suited for classifying datasets with non-linear relationships. Gaussian kernels, being a common and effective choice for SVMs, offer several advantages. They have the potential to improve accuracy for non-linearly separable data by modeling complex relationships. Furthermore, they are less prone to overfitting compared to some other kernels, a crucial consideration when dealing with limited data. The smooth decision boundaries generated by Gaussian kernels enhance generalization, leading to better performance on unseen data and resulting in more reliable and accurate predictions on new examples.

If the objective function for the SVM is convex with respect to the hyperparameters, Bayesian optimization guarantees finding the global minimum, leading to the best possible model performance. This differs from other optimizers that might get stuck in local optima. For convex objective functions, Bayesian optimization can converge to the optimal hyperparameters significantly faster compared to other methods. This is because it efficiently navigates the smooth landscape of the objective function, avoiding unnecessary exploration of irrelevant regions. This allows for a better understanding of how the SVM works and how its behavior changes with different hyperparameter settings which have been explained in this study later.

The Methods have been used to prepare effective feature engendering algorithm are:

1. **Differential Feature Extraction:** Taking the Gradient of Predictors when there are dependencies among them. The outcome provides lower dimensional dataset with specific differences by calculating the rate of changes of one or more predictors with respect to another. In this case, the Partial Differentiation of arithmetic average of axial values were taken for different frequencies.
2. **Wavelet Transformation and Denoising:** Performing Wavelet denoising on the Differential Features removes all the unnecessary signals from all the samples equally and increase the differences among the classes by removing similar properties.
3. **Visualization:** Visualization of Spectrums, Time-Frequency Domains, Power of Signals etc can visualize the differences among the classes. In this case the Power Spectrum Density plot visuality indicates the uniqueness of individual class and differences among classes.

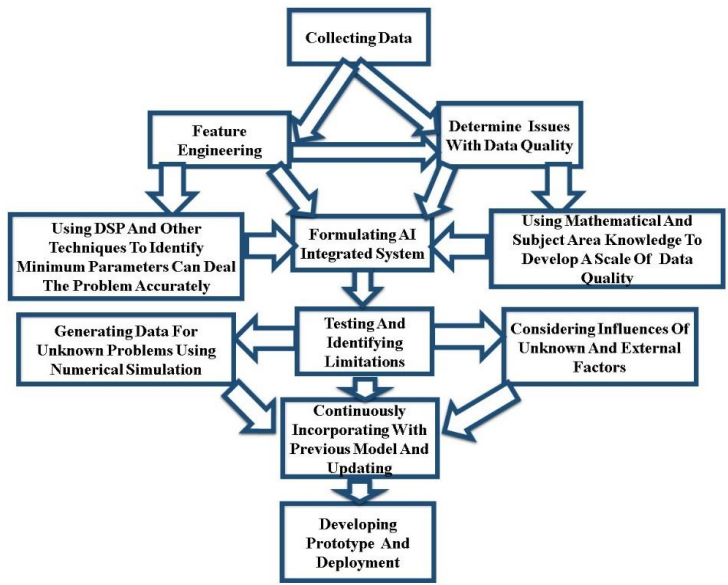

**Figure: 1: Visualization of the proposed research at a glance**

Various issues arise in rotating machines, each with different causes. Swiftly identifying and explaining the problem accurately is crucial. This work aims to uncover the causes of faults in rotating machines. To achieve this, data were collected using a specific procedure, setting up machines with different faults and obtaining measurements for each fault type. The data collection process is detailed in the data collection section. Promptly identifying and solving problems posed a challenge, leading to the deployment of a machine learning algorithm to predict faults automatically using collected data. All data were processed using innovative methods to classify and identify patterns of different faults quickly and accurately. Figures and tables visualize all processes, accompanied by detailed explanations. The methodology's summary is outlined below:

# 3 Data Collection and Feature Extraction

Data collection system and process of our previous published paper has been developed a little here and we're presenting the necessary parts along with the previous some discussion here [14].

## 3.1 Data Processing, Visualization and Explanations

Our goal was to extract valuable information into a single column for each experiment, accurately representing the inherent nature of the data. To achieve this, we performed various statistical and mathematical operations, experimenting with different methods. The introduced approach yielded highly accurate outcomes [4], [14], [19]. The different properties of the collected data are given shortly:

Initially, the average of columns containing axis values at different frequencies was computed. Subsequently, the dataset was formed by calculating the partial derivative of the averages concerning frequency, representing changes in data magnitude relative to different frequencies. The data, processed with machine learning models, demonstrated that varying magnitudes at different frequencies were produced by different faults, exhibiting distinct patterns. Differing magnitudes at different frequencies, contributing to distinct overall patterns, were manifested by structural looseness, misalignment, and bearing problems [19].

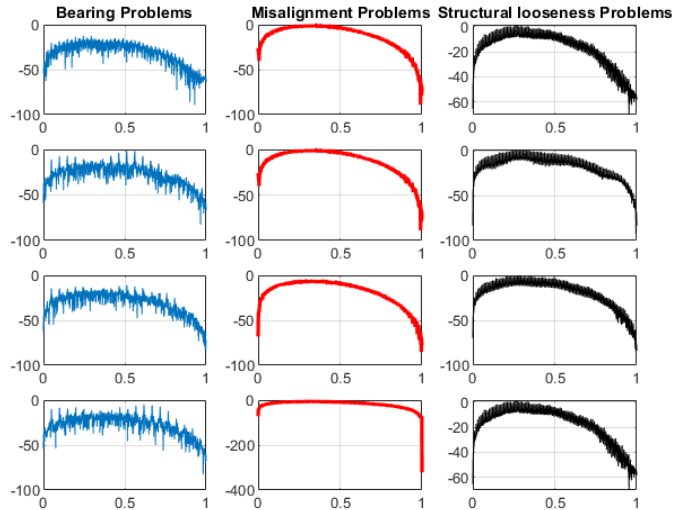

**Figure 10: Comparison among the patterns of three types features where X-axis indicates the Normalized frequency and Y-axis represents the Power of the signal in decibels**

### 3.1.1 SVM Classifier Configuration

Figures depicted the hyperparameter search history for both SVM and Neural Network models. The optimal hyperparameters, as indicated by Figure 14, include a one-versus-all approach, a box constraints value near 300, a kernel scaling parameter of 100, and unnecessary data standardization. The success of this hyperparameter combination reaffirms the accuracy of our data acquisition and feature engineering methods. The objective function evaluations, illustrated in Figure 14, are presented in a T-by-D table, where T is the number of evaluation points, and D is the number of variables. The multiclass method, box constraint, kernel scale, and standardized data serve as key indicators of the model's performance.

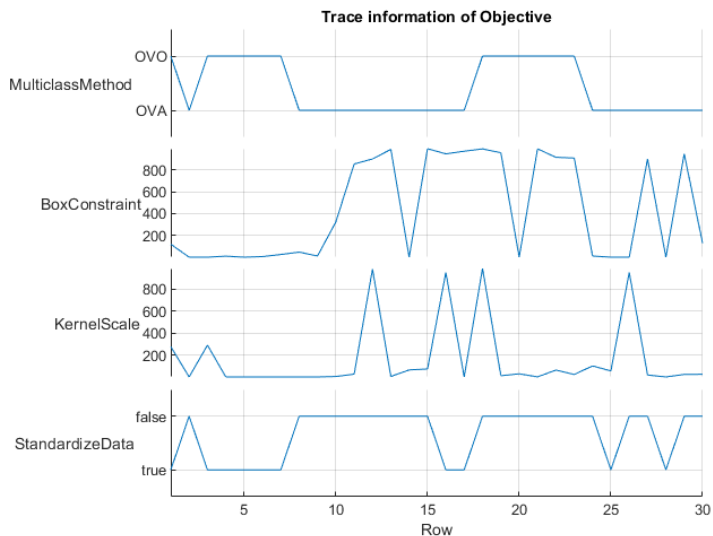

**Figure 14: Objective trace information of the SVM classification Model. The Figure shows different properties of an SVM classifier and indicating the change of one for another.**

From Figure 14, the best hyper parameter we add is one versus all; the box constant value was near 300, the kernel scaling parameter was found to be 100, and data standardization was not necessary. The outcome of the best combination of hyper parameters again confirms the correctness of our data acquisition and feature engineering method. From Figure 14, the points where the objective function was evaluated are specified as a T-by-D table, where T is the number of evaluation points, and D is the number of variables. Here, the multiclass method, box constraint, kernel scale, and standardize data are the key indicators of the model's performance.

**Table 1: Configuration of SVM Classifier**

|  | **Binary Learner's Property** | | |
|---|---|---|---|
| Parameter | Structural Looseness | Misalignment | Bearing Problem |
| Total Data Points | 42 | 118 | 80 |
| Support Vector Points | 16 | 41 | 44 |
| Bias | -3.10 | 1.034 | -0.5492 |
|  | **Model Properties** | | |
| Kernel Function | Gaussian | | |
| Binary Loss | Hinge | | |
| Kernel Scale | 23.12 | | |
| Minimum Objective | 0.0178699 | | |
| Optimizer | Bayesian Optimization | | |
| Model Solver | SMO (Sequential Minimum Optimizer) | | |

SVMs play a pivotal role in machine learning, particularly in binary classification tasks. They define the position of the hyperplane and the margin between classes, where the bias determines the hyperplane's relative location to data points. The minimum objective value represents the optimal solution to the optimization problem, rendering SVMs highly valuable. The primary goal is to find a hyperplane that maximizes the margin, with support vector points crucial in delineating class boundaries. Notably, SVMs excel in multiclass classification, utilizing methods like one-vs-one or one-vs-all for multiple classes. SVM classifier also adjusts elements in the predictor matrix through the Kernel Scale. The minimum of the objective function, vital for optimization, varies with hyperparameters such as box constraint and kernel scale [5], [28].

### 3.1.2 Neural Network Model Configuration

For our dataset a Neural Network with one input, two fully connected, and one output layer have selected as the optimal configuration [17]. The NN comprises two hidden layers, one input layer, and one output layer, utilizing the RELU activation function for the hidden layers and the SoftMax activation function for the output layer—SoftMax being a popular choice for multi-class classification. The input consists of the app's NN components, and the output represents three different faults. Appropriate hyperparameters were then determined to train the dataset and achieve the best outcome [3].

### 3.1.3 Convergence Information of Hyperparameters Optimization

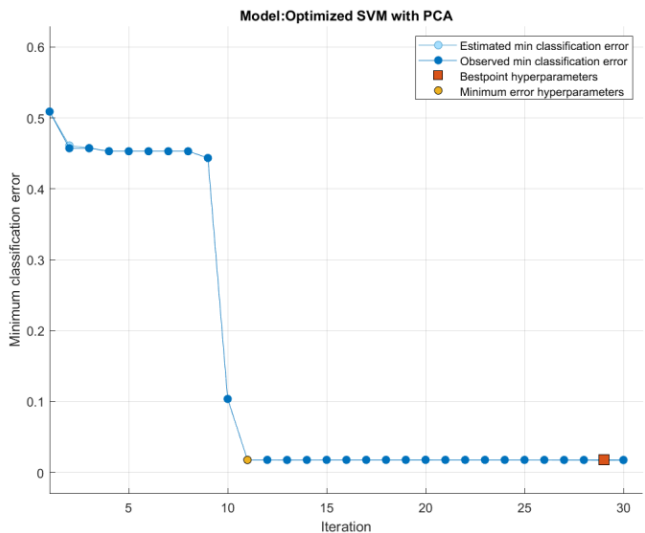

**Figure 17: Convergence information of the SVM model's hyperparameters search using Bayesian Optimization**

Figure 17 illustrates that the estimated and observed error for the SVM model are consistent. However, variations in the model's performance with different combinations of hyperparameters in terms of error are observed. Nevertheless, the model exhibits consistency in observed and estimated error minimization. The model shows good performance as indicated by the low minimum classification error achieved after about 10 iterations. There is consistency in the observed and estimated errors, showing that the model's performance is reliable. After an initial rapid decline in error, the model stabilizes quickly, indicating it has found an optimal solution efficiently.

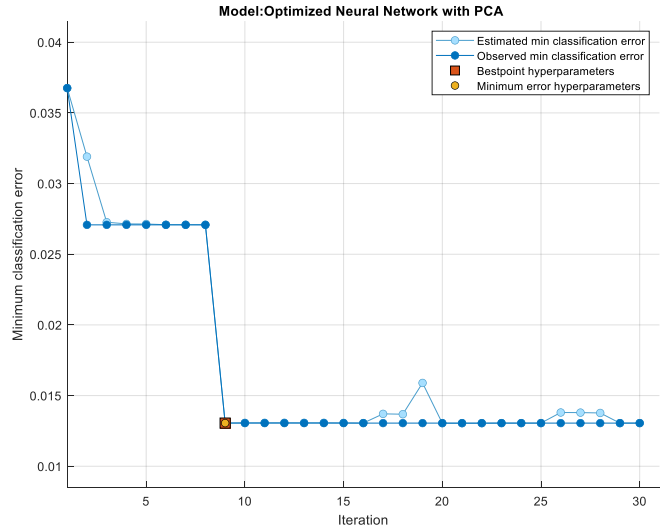

**Figure 18: Convergence information of NN model's hyperparameters search**

Nevertheless, the model exhibits consistency in observed and estimated error minimization. Figure 18 demonstrates that the NN model quickly identifies the combination of hyperparameters leading to minimum error and maximum accuracy simultaneously. However, we note that the observed and estimated error curves differ, and after the 15th iteration, the model overfits. Figures 17-18 highlight that the SVM classifier minimizes errors at a higher rate, stabilizing without fluctuations. In contrast, the NN classifier minimizes a small amount of loss within a short time, but the estimated minimum classification error does not converge at all.

## 3.2 Description of the Confusion Matrix

Confusion Matrix provide a deep insight into the overall training and prediction abilities of any classification model. Additionally, True positive and False negative rates are crucial in case of fault or damage identification because it's very sensitive in some cases like medical diagnosis and anything related to the human injury. The Confusion Matrix with the True Positive and False Negative Rates have been visualized and described in this section.

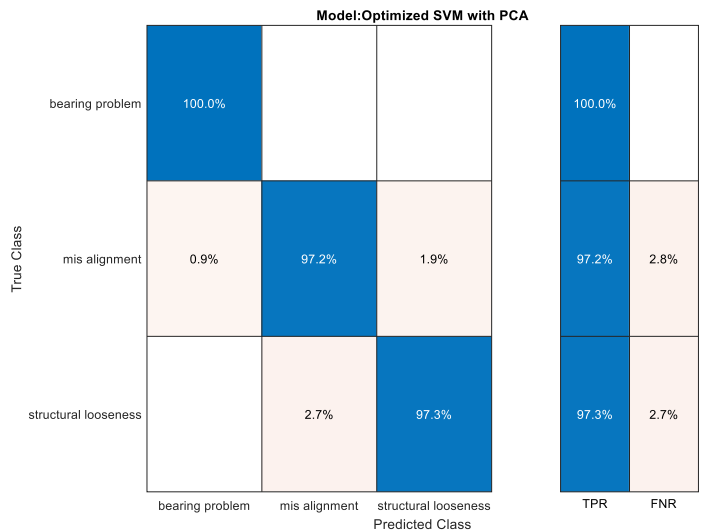

**Figure 19: Confusion matrix of SVM model representing the predictions on the validation data with True Positive and False Negative rate**

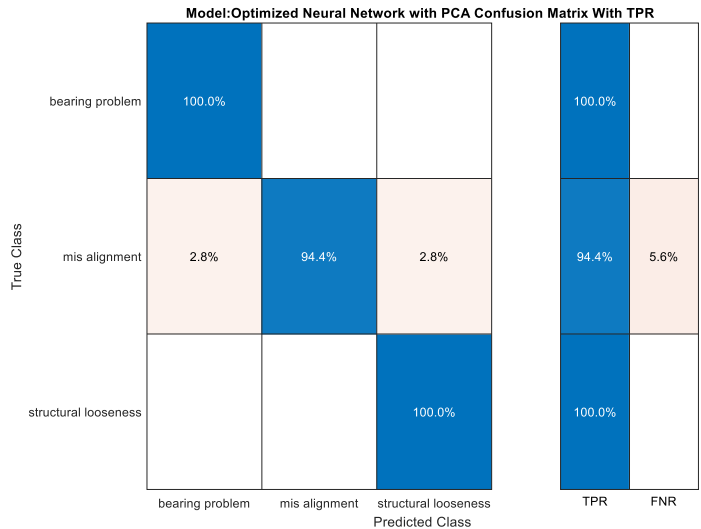

**Figure 20:** Confusion matrix of Neural Network model representing the predictions on the validation data with True Positive and False Negative rate

The Confusion Matrix in Figure 19 illustrates stable and effective training. It successfully predicted all validation data for the bearing problem due to an ample number of samples to identify the pattern. The SVM model predicted with minimal error in the other two cases of Misalignment and structural looseness. Figures 19-20 clearly indicate that the NN model performs well and accurately predicts the validation dataset for bearing problems and structural looseness. However, in the case of the misalignment problem, the NN classifier achieved an accuracy of only 94%. On the other hand, the SVM classifier predicted the bearing problem with 100% accuracy and the misalignment and structural looseness problems with 97.2% and 97.3% accuracy. This underscores that the SVM classifier performs well with both fewer and larger training samples. Given ample samples in bearing and structural looseness but fewer samples in misalignment, the NN classifier perfectly predicted the patterns of bearing and structural looseness but struggled to maintain accuracy in the case of misalignment. In contrast, SVM predicted all classes effectively, maintaining consistency and alignment with the training samples per class, indicating SVM's capability to capture the underlying pattern of the dataset on average.

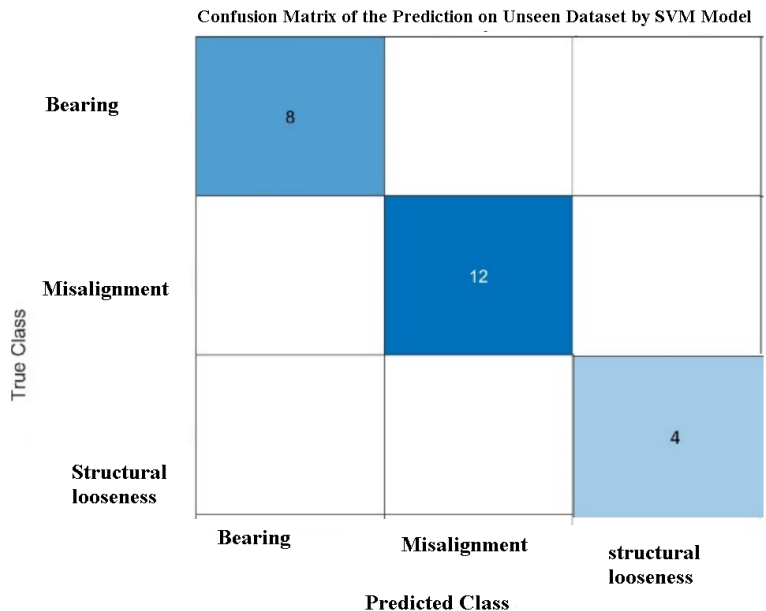

**Figure 21:** Prediction on test dataset with SVM classifier, the confusion matrix shows all the three classes were identified correctly

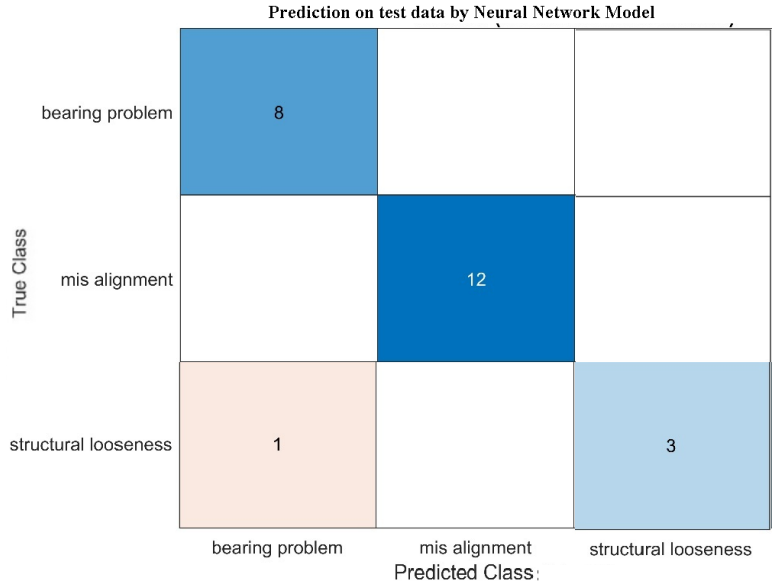

**Figure 22:** Prediction on test dataset with Neural Network classifier, the confusion matrix shows all 12 samples of Misalignment problems were identified correctly and the model failed to identify the pattern of bearing problem and structural loosens as the model identified one structural loosens sample's signal as bearing problem.

The loss function can be modified to improve the outcome of NN classification. Additionally, we utilized the rectified linear unit as the activation function in the hidden layer, but exploring other activation functions can provide insights into the performance of NNs. The NN classifier employed a structure with 25 neurons in each layer and utilized the LBFGS optimization algorithm to efficiently find local minimum values for its objective function. Notably, this configuration achieved an impressive accuracy of over 97% when predicting new data, showcasing its potential for faster and more accurate predictions with increased data volume. While the performance of the NN was not optimal, it still achieved satisfactory results, boasting a 97.2% accuracy using principal component analysis on limited data. This underscores the effectiveness of data processing and feature extraction methods. We illustrate the behaviour of the bias over iterations, revealing a quadratic curve for specific layers, which is easily minimized [4], [13].

### 3.3 Assessment and Comparison of SVM and Neural Network Classifier

In this stage, the summary of machine learning model's performance has been evaluated and discussed with the tables and necessary visualizations.

**Table 2: Comparison Between the SVM and Neural Network Classifier**

|  | SVM Classifier Model | Neural Network Classifier Model | Remarks |
|---|---|---|---|
| Accuracy (Validation) | 98.2% | 97.1% | Indicates Accuracy of prediction for Validation dataset. |
| Accuracy (Test data 20% of Training) | 99.5% | 92.3% | Indicates Accuracy of prediction on new data or samples. |
| Accuracy (Test data Random) | 100% | 89.2% | Indicates Accuracy of prediction on new data or samples. |
| Training Time | 257 Seconds | 696 Seconds | Lesser means Reliable, interpretable, and appropriate for larger data and commercial applications. |
| Iterations | 30 | 18 | Faster training indicates the power of capturing the pattern or how the model fits well with the dataset. |
| Prediction Speed | 130 orbs/seconds | 120 orbs/seconds | Faster indicates appropriate for real-time and emergency situations. |
| Total Number of Principle Components (Predictors) | 18 | 18 | Indicates that the dataset is processed accurately and efficiently. |
| Error minimization performance | Faster and stable. | Slower and unstable. | Faster means accurate and efficient, and the dataset is appropriately prepared. |

The objective trace information represents the effectiveness of hyperparameters optimization algorithms. Since this study employed the Bayesian Optimization algorithm to find optimal hyperparameters that can generate the best and stable predictions on test data, it is necessary to explain and visualize the effectiveness of our selection to justify the choice of the hyperparameter Optimization Algorithm we previously described. Figures 13 and 15 illustrate how fast and accurately the hyperparameters such as constraints, kernel scaling value, and multi-class methods have been revealed in the case of Neural Network and SVM. Although we find different information regarding the objective and the objective functional in the case of the SVM and NN classifier, we can get an idea about the effectiveness and accuracy of the parameters by observing the consistency in the convergence of every hyperparameter. In the case of the Support Vector Machine Classifier, we can clearly see that all the parameters follow the same pattern, like the change in the box constraint values, kernel scaling values, and standardized data method follow the same pattern with respect to its iteration. But in the case of the NN Classifier, we can see that the loss function and gradient were minimized at a very fast rate, but the time taken for iteration and the number of steps fluctuated since the last iteration. This type of inconsistency indicates the effectiveness of the hyperparameter searching algorithm and clearly states that the Support Vector Machine outperformed the NN classifier [29], [30].

# 4 Key Findings

Key observations on the performance and effectiveness of the process, encompassing feature extraction and model building:

**Feature Extraction:**

- **Effective feature identification:** The extracted features successfully capture distinctive characteristics of each fault type, enabling accurate classification.
- **Frequency domain relevance:** Power spectrum density estimation proves valuable for fault diagnosis in this domain, revealing characteristic patterns associated with different fault types.
- **Potential for further refinement:** Exploring additional features or feature combinations might enhance classification performance even further.

**Model Building:**

- **SVM's overall superiority:** SVM outperformed the neural network in terms of accuracy, generalization, training time, and robustness to outliers, making it the preferred choice for this task.
- **Neural network's potential:** While not as successful in this specific case, neural networks hold promise for complex problems or larger datasets, warranting further exploration with careful hyperparameter tuning.

**Key Performance Highlights:**

- **Excellent accuracy:** Both models achieved high accuracy on the training set, with SVM maintaining perfect accuracy on the unseen test set.
- **SVM's generalization strength:** SVM demonstrated superior generalization ability, accurately classifying all unseen samples, while the neural network misclassified one.
- **Efficient training:** SVM trained significantly faster than the neural network (257 seconds vs. 696 seconds).

**Recommendations:**

- **Prioritizing SVM for this problem:** SVM's overall performance renders it the recommended choice for this fault diagnosis task.
- **Considering feature expansion:** Investigate incorporating additional features or feature combinations to potentially improve classification accuracy.
- **Addressing neural network's overfitting:** Mitigate the neural network's tendency to overfit through regularization techniques or data augmentation.

The process from feature extraction to model building has been successful in developing effective fault diagnosis models. SVM emerged as the superior choice for this problem, demonstrating excellent accuracy, generalization, and training efficiency. Continuous evaluation and refinement of features and model choices can further enhance the diagnostic capabilities.

# 5 Conclusion

In conclusion, this research has successfully developed a cost-effective data acquisition system using ESP32. The study highlights the shortcomings of conventional fault detection and diagnosis systems in limited-resource situations and proposes the implementation of a data acquisition system with hardware that is affordable for all industries in the Least Developed Country (LDC) context. To overcome the significant challenge of making successful predictions with limited and noisy data, a combination of mathematical and statistical techniques, such as wavelet analysis, a new concept of differential feature extraction, and principal component analysis, was implemented sequentially with an SVM classifier for predictions.

The configuration of the prediction model was determined using mathematics and concepts from dynamical system analysis and functional analysis to avoid the trial-and-error method. Despite limited data and resources, state-of-the-art methods like deep learning and transfer learning failed to perform well, especially in the situation of least developed countries. The need to reconsider existing systems of automatic Fault Detection and Diagnosis (FDD) is crucial to ensure the occupational health and safety of all. While some studies have developed new optimization algorithms for this purpose, others have focused on implementation, often overlooking theoretical aspects. In this study, we have explained the theoretical and experimental aspects and applications in a manner understandable by academics and individuals with an industrial background. Rather than using rigorous mathematical terms and equations, we aimed to keep the content graphical and described it with more extended text. We incorporated well-known methods and models in this study so that anyone can comprehend and learn the concepts and methods easily and extend or modify them according to their needs.

# 6 References

[1] S. L. Nystrom and D. J. McKay, "Memes: A motif analysis environment in R using tools from the MEME Suite," *PLoS Comput Biol*, vol. 17, no. 9, p. e1008991, Dec. 2021, doi: 10.1371/journal.pcbi.1008991.

[2] H. Liu, R. Ma, D. Li, L. Yan, and Z. Ma, "Machinery Fault Diagnosis Based on Deep Learning for Time Series Analysis and Knowledge Graphs," *J Signal Process Syst*, vol. 93, no. 12, pp. 1433–1455, Dec. 2021, doi: 10.1007/s11265-021-01718-3.

[3] D. Leake, Z. Wilkerson, V. Vats, K. Acharya, and D. Crandall, "Examining the Impact of Network Architecture on Extracted Feature Quality for CBR," pp. 3–18, 2023, doi: 10.1007/978-3-031-40177-0_1.

[4] F. M. Shakiba, S. M. Azizi, M. Zhou, and A. Abusorrah, "Application of machine learning methods in fault detection and classification of power transmission lines: a survey," *Artif Intell Rev*, vol. 56, no. 7, pp. 5799–5836, Dec. 2023, doi: 10.1007/S10462-022-10296-0.

[5] A. Abid, M. T. Khan, and J. Iqbal, "A review on fault detection and diagnosis techniques: basics and beyond," *Artif Intell Rev*, vol. 54, no. 5, pp. 3639–3664, Dec. 2021, doi: 10.1007/S10462-020-09934-2.

[6] Y.-J. Park, S.-K. S. Fan, and C.-Y. Hsu, "A Review on Fault Detection and Process Diagnostics in Industrial Processes," *Processes*, vol. 8, no. 9, p. 1123, Dec. 2020, doi: 10.3390/pr8091123.

[7] L. Ming and J. Zhao, "Review on chemical process fault detection and diagnosis," in *2017 6th International Symposium on Advanced Control of Industrial Processes (AdCONIP)*, IEEE, Dec. 2017, pp. 457–462. doi: 10.1109/ADCONIP.2017.7983824.

[8] X. Du, "Fault detection using bispectral features and one-class classifiers," *J Process Control*, vol. 83, pp. 1–10, Dec. 2019, doi: 10.1016/j.jprocont.2019.08.007.

[9] B. E. Goodlin, D. S. Boning, H. H. Sawin, and B. M. Wise, "Simultaneous Fault Detection and Classification for Semiconductor Manufacturing Tools," *J Electrochem Soc*, vol. 150, no. 12, p. G778, 2003, doi: 10.1149/1.1623772.

[10] I. Hwang, S. Kim, Y. Kim, and C. E. Seah, "A Survey of Fault Detection, Isolation, and Reconfiguration Methods," *IEEE Transactions on Control Systems Technology*, vol. 18, no. 3, pp. 636–653, Dec. 2010, doi: 10.1109/TCST.2009.2026285.

[11] "Department of Inspection for Factories and Establishments Ministry of Labour and Employment National Profile on Occupational Safety and Health in Bangladesh 2019." [Online]. Available: www.dife.gov.bd

[12] I. Rahman, "A Review on Occupational Health Safety in Bangladesh with Respect to Asian Continent." 2016.

[13] M. Fernandes, J. M. Corchado, and G. Marreiros, "Machine learning techniques applied to mechanical fault diagnosis and fault prognosis in the context of real industrial manufacturing use-cases: a systematic literature review," *Applied Intelligence*, vol. 52, no. 12, 2022, doi: 10.1007/s10489-022-03344-3.

[14] A. H. M. Ripon and M. A. Ullah, "Rotating Machine Fault Detection Using Support Vector Machine (SVM) Classifier," *2023 4th International Conference on Computing and Communication Systems, I3CS 2023*, 2023, doi: 10.1109/I3CS58314.2023.10127320.

[15] D. Gonzalez-Jimenez, J. del-Olmo, J. Poza, F. Garramiola, and P. Madina, "Data-Driven Fault Diagnosis for Electric Drives: A Review," *Sensors*, vol. 21, no. 12, p. 4024, Dec. 2021, doi: 10.3390/s21124024.

[16] N. M. Nor, C. R. C. Hassan, and M. A. Hussain, "A review of data-driven fault detection and diagnosis methods: applications in chemical process systems," *Reviews in Chemical Engineering*, vol. 36, no. 4, pp. 513–553, Dec. 2020, doi: 10.1515/revce-2017-0069.

[17] Y. Huang, C. H. Chen, and C. J. Huang, "Motor fault detection and feature extraction using rnn-based variational autoencoder," *IEEE Access*, vol. 7, pp. 139086–139096, 2019, doi: 10.1109/ACCESS.2019.2940769.

[18] M. A. Trovero and M. J. Leonard, "Time Series Feature Extraction." pp. 2018–2020.

[19] A. Entezami, "Feature Extraction in Time Domain for Stationary Data," 2021, pp. 17–45. doi: 10.1007/978-3-030-66259-2_2.

[20] "https://en.wikipedia.org/wiki/Dynamic_time_warping."

[21] P. Kumar and A. S. Hati, "Review on Machine Learning Algorithm Based Fault Detection in Induction Motors," *Archives of Computational Methods in Engineering*, vol. 28, no. 3, pp. 1929–1940, Dec. 2021, doi: 10.1007/S11831-020-09446-W.

[22] J. Yu and Y. Zhang, "Challenges and opportunities of deep learning-based process fault detection and diagnosis: a review," *Neural Comput Appl*, vol. 35, no. 1, pp. 211–252, Dec. 2023, doi: 10.1007/S00521-022-08017-3.

[23] S. Y. Shao, W. J. Sun, R. Q. Yan, P. Wang, and R. X. Gao, "A Deep Learning Approach for Fault Diagnosis of Induction Motors in Manufacturing," *Chinese Journal of Mechanical Engineering (English Edition)*, vol. 30, no. 6, pp. 1347–1356, Dec. 2017, doi: 10.1007/S10033-017-0189-Y.

[24] B. Vincent, C. Duhamel, L. Ren, and N. Tchernev, "A PCA and SVR based method for continuous industrial process modelling," *IFAC-PapersOnLine*, vol. 51, no. 11, pp. 1604–1609, 2018, doi: 10.1016/j.ifacol.2018.08.264.

[25] X. Jin, Y. Zhang, and D. Yao, "Simultaneously Prediction of Network Traffic Flow Based on PCA-SVR," in *Advances in Neural Networks – ISNN 2007*, Berlin, Heidelberg: Springer Berlin Heidelberg, pp. 1022–1031. doi: 10.1007/978-3-540-72393-6_121.

[26]  J. Tang, C. Fu, C. Mi, and H. Liu, "An interval sequential linear programming for nonlinear robust optimization problems," *Appl Math Model*, vol. 107, pp. 256–274, Jul. 2022, doi: 10.1016/J.APM.2022.02.037.

[27]  J. Tang *et al.*, "A possibility-based solution framework for interval uncertainty-based design optimization," *Appl Math Model*, vol. 125, pp. 649–667, Jan. 2024, doi: 10.1016/J.APM.2023.09.010.

[28]  J. Chen, B. Xu, and X. Zhang, "A Vibration Feature Extraction Method Based on Time-Domain Dimensional Parameters and Mahalanobis Distance," *Math Probl Eng*, vol. 2021, pp. 1–12, Dec. 2021, doi: 10.1155/2021/2498178.

[29]  A. M. Elshewey, M. Y. Shams, N. El-Rashidy, A. M. Elhady, S. M. Shohieb, and Z. Tarek, "Bayesian Optimization with Support Vector Machine Model for Parkinson Disease Classification," *Sensors*, vol. 23, no. 4, p. 2085, Feb. 2023, doi: 10.3390/s23042085.

[30]  R. Turner *et al.*, "Bayesian Optimization is Superior to Random Search for Machine Learning Hyperparameter Tuning: Analysis of the Black-Box Optimization Challenge 2020." [Online]. Available: http://clopinet.com/isabelle/Projects/NIPS2006/

[31]  Paul, Arindam Kumar; Md. Ripon, Abu Hanif (2024), "Design & Implementation of Automatic Machine Condition Monitoring and Maintenance System in Limited Resource Situations", Mendeley Data, V1, doi: 10.17632/6kxbbjjp7g.1